# Theoretical model of external spinal cord stimulation

**Mikhail N. Shneider [1], Mikhail Pekker[2]**

[1]Department of Mechanical and Aerospace Engineering, Princeton University, Princeton, NJ USA
[2]Retired

*m.n.shneider@gmail.com

**Abstract**
In this paper, a simple theoretical model of excitation of action potentials of multiple motor pools by stimulating current pulses over the lumbosacral regions of the spinal cord is presented. The present model is consistent with known experimental data.

## I.  Introduction

In the last decade, a fairly large number of experimental studies have appeared in which the effectiveness of restoring control over the lower limbs during spinal stimulation of the lumbosacral regions of the spinal cord below the site of the spinal cord injury has been shown. These effects were observed with the use of electrical stimulation using implanted epidural electrodes and with stimulation using non-invasive electrodes placed directly on the skin over the lower spinal cord [1-9].

Computer model for epidural electrical stimulation of spinal nerve has two parts: model for calculating current system in region of spinal nerve and a model of the excitation of action potential in axon fiber by these currents [10-18]. The calculation of currents in these works was carried out on the basis of solving the continuity equation for currents:

$$\nabla \cdot \vec{j} = \nabla \cdot \sigma \vec{E}_e = \nabla \cdot \sigma \nabla \varphi_e = 0, \qquad (1)$$

with Dirichlet boundary conditions set at the outermost boundaries of the model:

$$\varphi_e(d\Omega) = 0 \qquad (2)$$

where $\varphi, \vec{j}, \vec{E}$ are distributions of potential, currents and electric field in electrolyte, $\sigma$ is the conductivity, $d\Omega$ is the outermost surface (saline) of the model. It was assumed in this model that the source of the current was a point electrode located outside the spinal cord. In fact, boundary condition (2) corresponds to an electrode, and has nothing to do with the presence of a conductive fluid in which the spinal nerve is located. A correct numerical model for calculating the distribution of currents in the spinal nerve region (equation (1)) should include a second electrode.

The model of action potential initiation by currents flowing through the spinal cord in the axons of motor myelinated neurons, described in [11, 17, 18], has the form:

$$\frac{d\varphi_n}{dt} = \frac{1}{C_m}\left(G_a(\varphi_{i,n+1} - 2\varphi_{i,n} + \varphi_{i,n-1}) + (\varphi_{e,n+1} - 2\varphi_{e,n} + \varphi_{e,n-1}) - I_{i,n}\right) \qquad (3)$$

$$\varphi_n = \varphi_{i,n} - \varphi_{e,n} + \varphi_r \qquad (4)$$

Were $I_{i,n}$ is the ionic current passing through the nth active part of the membrane with capacity $C_m$ and conductance $G_a$, $I_{i,n}$ is a function of the voltage between the inside potential $\varphi_{i,n}$ and outside potential $\varphi_{e,n}$ в узле n, $\varphi_r$ is the inner resting potential. From (3) it follows that if the external potential on the membrane, determined by the currents flowing in the spinal cord, is constant, then the initiation of the action potential will not occur. This is not true, since the initiation of the action potential occurs due to the charging of the axon membrane by currents in the spinal cord, initiated by electrodes (the axon membrane is a cylindrical condenser filled with saline). An increase in the potential difference across the membrane lowers the excitation threshold of the action potential.

The purpose of this work is development of a simple physically adequate model explaining the known experimental results. Our physical model is based on a very simple idea: the currents generated by external current sources [19, 20] charge the neuron membranes potential to the level at which the action potential is initiated. Since the potential on the non-conducting membrane $\varphi_a$ is proportional to the radius of the ax on $a$, at the same current density in the extracellular fluid and the same conductivity of the extracellular medium, the change in potential difference across the membrane will be different for neurons with axons of different radii. Accordingly, with the same threshold potential difference, the excitation of neurons with axons of a larger radius should occur at a lower current density than neurons with axons of small radius.

## II. Description of the experiment

Let us briefly describe the formulation of a very indicative experiment [9]. Conductive rubber electrode with a radius $R_{el} = 9$ mm placed on the skin between the spinous processes T12-L1, T11-T12, T10-T11 as a cathodes, and two $5 \times 9$ cm self-adhesive electrodes (Pro-Patch) located symmetrically on the skin over the iliac crests as anodes. Stimulation was delivered as single square-wave pulse every 6s. The stimulation intensity was varied from $I_0 = 2$ mA till 100 mA [6] Fig.1. For electrical stimulation was used monopolar rectangular stimuli (1 ms duration) filled with a carrier frequency of 10 kHz and at an intensity ranging from 80 to 180 mA (30 Hz at T11 and 5 Hz at Co1).

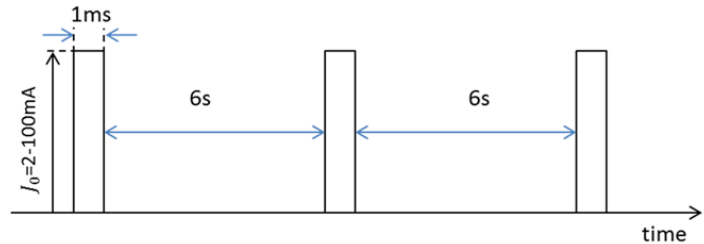

**Figure 1.** Periodical current pulses applied to the electrodes.

## III. Theoretical model

Since the radius of the spinal nerve does not exceed 2.4 mm [21], and the distance from the nerve fiber (spinal tissue) to the electrode (cathode) (Fig. 2 A) is certainly less than the electrode radius ~ 5 mm, the distribution of currents in the region of spinal cord can be considered perpendicular to the electrode and uniformly distributed over the area far from the spinal nerve (Fig.2B).

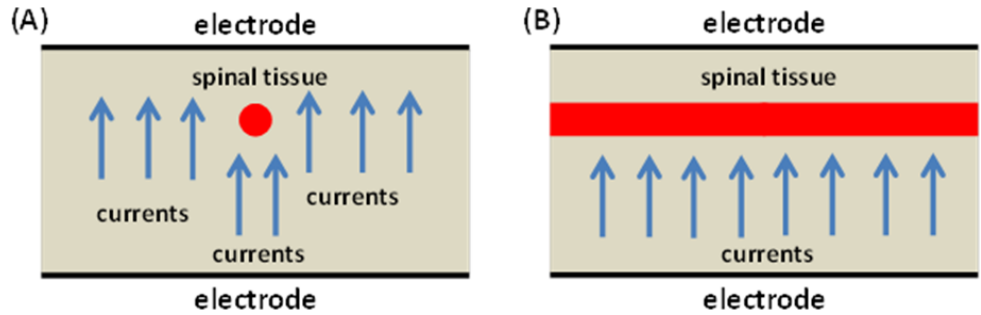

**Figure 2**. A sketch of the location of the spinal tissue in the area of current flowing between the electrodes. (A) – cross section of spinal cord; the distance from spinal nerve is much less than the radius of the electrode. (B) – longitudinal section of spinal cord; the size of the area of charging of the surface of neurons is much larger than the spinal tissue radius (Fig 3).

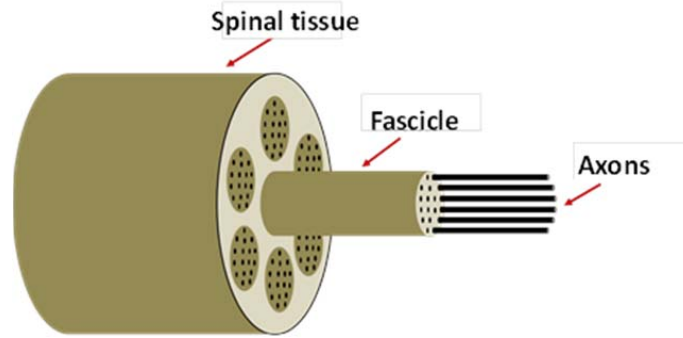

**Figure 3.** Schematic of a spinal tissue depicted as a bundle of isolated wires (fascicles), consisting of thin wires (axons).

Consider the following model, based on conditions, corresponding to Fig. 2 and 3:

1. The axons of the fascicles are thin-walled cylinders that are far enough from each other so that they can be viewed independently of each other in the field of constant currents (Fig. 4).
2. The axon environment is an electrolyte with conductivity of the order of 2-5 S/m.
3. The internal environment of the axon (cylinder) is also an electrolyte with conductivity close to the conductivity of the external medium.
4. The membrane is impermeable to ionic conduction currents, which charge the capacitance, i.e. the current in the electrolyte is closed through the membrane capacitance by the displacement current, $c_m dU_m/dt$, where $c_m$ the membrane capacitance per unit area, $U_m$ is the voltage on the membrane. The axon (cylinder) membrane is charged with currents in the electrolyte until the additional surface charge accumulating on the membrane compensates for the radial field of the currents that charge the membrane.

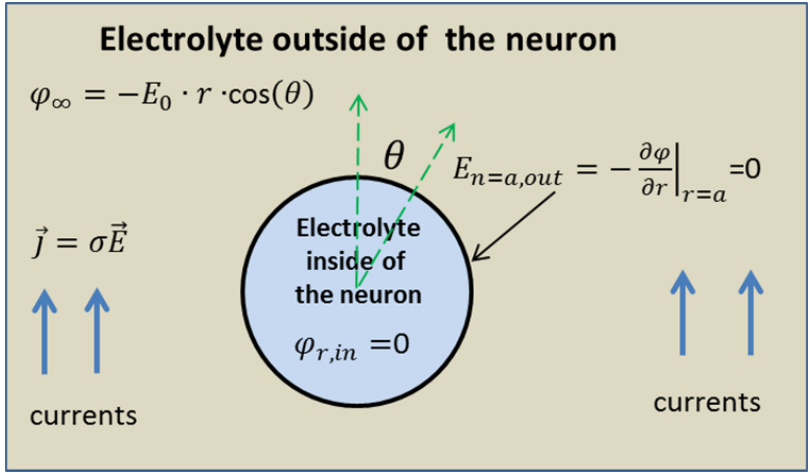

**Figure 4.** Schematic of a solitary axon cross-section in a field of currents uniformly distributed over the volume.

Consider the case when the time of the field change in the electrolyte is much longer than the charging time of the surface of the cylinder (the neuron membrane), shown in Fig. 1. Eation (1) for the currents flowing in the external electrolyte has the form:

$$div\vec{j} = -\sigma \left(\frac{1}{r}\frac{\partial}{\partial r}r\frac{\partial \varphi}{\partial r} + \frac{1}{r^2}\frac{\partial^2 \varphi}{\partial \theta^2}\right) = 0, \qquad (5)$$

where it is assumed that the conductivity $\sigma$ of the electrolyte is constant.

Far from the cylinder, the current density is constant, since the current is determined by the flat electrodes used in experiments [21], as shown schematically in Fig. 2. Accordingly, the current, potential and the electric field are related by the relation:

$$\varphi_\infty = -E_0 r cos(\theta) = -j_0 r cos(\theta)/\sigma, \qquad (6)$$

where $j_0 \approx I_0/\pi R_{el}^2$. Since the membrane is impermeable to ion currents, the radial electric field

$$E_n = -\frac{\partial \varphi}{\partial r}\bigg|_{r=a} \qquad (7)$$

must go to zero, as a result of charging. Hereafter, $a$ is the radius of the cylinder.
Since, when the stationary state is reached, the currents inside the conductive cylinder (axon) are absent, (the walls of the cylinder are impermeable to currents), the potential of the cylinder is uniform and constant. For definiteness, we set it equal to zero.

$$\varphi_{in} = 0 \qquad (8)$$

The solution of equation (5), taking into account the boundary condition (6), has the form:

$$\varphi = -E_0 r cos(\theta) + \sum_1^\infty B_n r^{-n} cos(n\theta) \qquad (9)$$

Since the radial field is zero at the boundary of the cylinder, from (9) we obtain:

$$E_n = -\frac{\partial \varphi}{\partial r}\bigg|_{r=a} = E_0 \cos(\theta) - \sum_1^\infty n B_n a^{-n-1} \cos(n\theta) = 0 \tag{10}$$

And

$$B_1 = -E_0 a^2, \quad B_2 = B_3 = \cdots = B_n = 0 \tag{11}$$
$$\varphi = -E_0 \cos(\theta)\left(r + \frac{a^2}{r}\right), \tag{12}$$

Respectively:

$$E_r = E_0 \cos(\theta)\left(1 - \frac{a^2}{r^2}\right) = \frac{j_0}{\sigma}\cos(\theta)\left(1 - \frac{a^2}{r^2}\right) \tag{13}$$
$$E_\theta = E_0 \sin(\theta)\left(1 + \frac{a^2}{r^2}\right) = \frac{j_0}{\sigma}\sin(\theta)\left(1 + \frac{a^2}{r^2}\right) \tag{14}$$

In accordance with (12) and (9), the potential difference on the membrane is:

$$\varphi_a = -2E_0 a \cdot \cos(\theta) = -\frac{2j_0}{\sigma} a \cdot \cos(\theta) = -\frac{2I_0}{\pi R_{el}^2 \sigma} a \cdot \cos(\theta) \tag{15}$$

It follows from (15) that the steady-state potential difference on the membrane depends only on the magnitude of the current and the radius of the cylinder. Fig. 5 shows the current lines in the vicinity of the cylindrical axon charged to a potential (15) which are corresponding to the electric field with components (13) and (14).

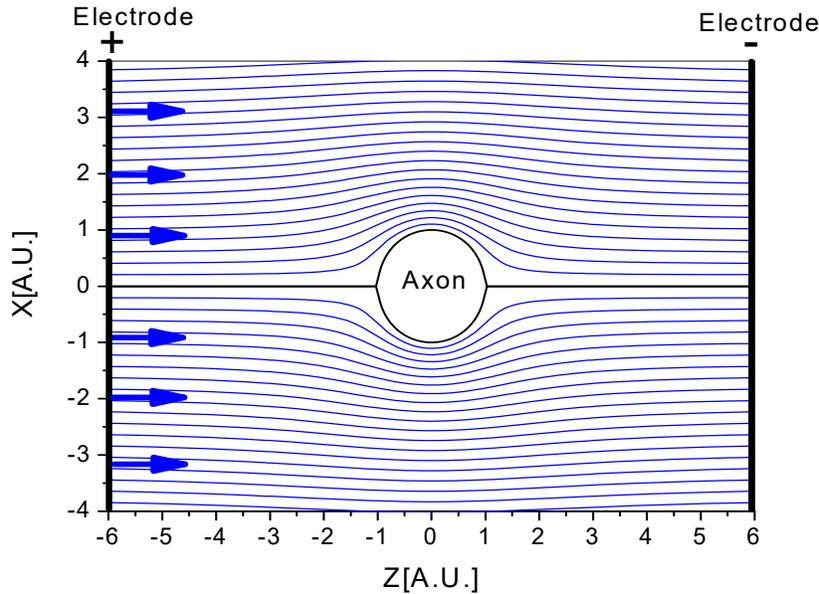

**Figure 5.** The current lines in the vicinity of the cylindrical axon charged to a potential (15) corresponding to the electric field with components (13) and (114).

It is seen that near the cylinder the field has only a tangential component. Now we answer the question how quickly the potential difference is established on the membrane when the current is switched on, as in the figure (1). Obviously, this time is determined by the time required to charge the cylindrical capacitor to a potential equal to (15). For simplicity, we shall assume that considered thin-walled dielectric cylinder does not perturb the electric field near its surface. In this case, the radial current charging the cylindrical capacitor is:

$$j_n = -j_0 \cos(\theta) \tag{16}$$

In this case, the estimate of the transition time (characteristic charging time of the capacitor) is:

$$\tau_{tr} \sim c_m \frac{\varphi_a}{j_n} = 2 \frac{c_m a}{\sigma}, \tag{17}$$

where $c_m \approx 10^{-2} \left[\frac{F}{m^2}\right]$ is the membrane capacitance per unit area [22]. Let us show that this time in experiments [5-9] is several orders of magnitude shorter than the duration of the current pulse in Fig. 1 (1ms) and, therefore can be neglected. According to [22], the conductivity of the electrolyte outside the neuron is $\sigma \sim$ 1-3 $\Omega^{-1}m^{-1}$, but in the Gray Matter region $\sigma \approx 0.25$ $\Omega^{-1}m^{-1}$ and in the White Matter $\sigma \approx 0.08$-0.6 $\Omega^{-1}m^{-1}$, where the spinal nerve is located [13-25]. The radius of motoneurons $a$ lies within 6-10 μm for α motor neurons, within 2.5-6 μm for β motoneurons, within 1.5-3 μm for γ neuron, and within 1-2.5 μm for δ neuron [26]. Table 1 presents estimates of the value of τ (Eq. 17) for different types of neurons obtained for different $a$ at $\sigma = 1$ $\Omega^{-1}m^{-1}$ and $\sigma \approx 0.25$ $\Omega^{-1}m^{-1}$.

**Table 1.** The transitional time τ (Eq. 13) to charge the surface of motor neurons of different types at different ambient conductivity [26].

| Type | Radius [μm] | Average [μm] | $\tau[\mu s]$ $\sigma = 1$ $\Omega^{-1}m^{-1}$ | $\tau[\mu s]$ $\sigma = 0.25$ $\Omega^{-1}m^{-1}$ |
|---|---|---|---|---|
| α | 6-10 | 8 | 0.160 | 0.64 |
| β | 2.5-6 | 4.25 | 0.085 | 0.340 |
| γ | 1.5-3 | 2.25 | 0.045 | 0.180 |
| δ | 1-2.5 | 1.75 | 0.035 | 0.140 |

## IV. Results and Discussion

Since in experiments [5-9] the duration of the current pulse was 1 ms, that is almost 4 orders of magnitude longer than the charging time of the neuron membrane, the potential difference on the membrane is constant throughout the duration of the current pulse and is given by formula (16). Substituting the value $E_0$ corresponding to the current at the electrode equal to 100mA, the radius of the electrode $R_{el} = 0.9 \cdot 10^{-2}$ m (10) and the radius of the motoneuron $a$=8μm, we get that for $\sigma = 1$ $\Omega^{-1}m^{-1}$, $U_m \approx 8$ mV, and for $\sigma = 0.25$ $\Omega^{-1}m^{-1}$, $U_m \approx 32$ mV. The obtained values of $U_m$ lie within the region of the potential perturbations at which the action potential starts.

In experiments [9], the current was a step function and varied within $I_0$ =2-100 mA (Fig. 1). Thus, at currents of around 10 mA, 20mA and 30 mA, the action potentials excitation was observed in motoneurons. This result can be easily explained. Indeed, since with the same initiating voltage $\varphi_a = U_m$ on the membrane, the critical current is inversely proportional to the radius of the axon (15):

$$I_0 \approx 2 \frac{\pi R_{el}^2 \sigma}{a} U_m \tag{18}$$

and the ratios of the mean radii of α, β and γ neurons are 1.9: 1: 0.53, then, at the same initiating voltage, the critical current has to be 10.5:20:38 mA for $U_m = 10 mV$, $R_{el} = 0.9 \cdot 10^{-2} m$, $\sigma = 0.061 [\Omega^{-1} m^{-1}]$. The calculated values of the critical currents for the formulated conditions are given in Table 2. Thus, the excitation of different types neurons occurs at different critical currents. For example, if the current exceeds the critical value enough only for the excitation of axons from the larger α-neurons, axons from other types of neurons would remain inactive within a fascicle of axons. It should be pointed out, however, that the dynamic environment and different cellular and synaptic components within and among spinal networks at varying current levels functioning under *in vivo* conditions is much more complex than that of a bundle of axons within a nerve fascicle. Intuitively, it would seem that this more complex environment under *in vivo* conditions would make the probability of excitation even higher than that when modeling such a phenomenon based on an axonal diameter.

The work [9] considered the influence of different currents applied transcutaneously at three different sites along the spinal axis T12-L1, T11-T12 and T10-T11. In the T12-L1 region, the excitation threshold was at a current of 15 mA, in the T11-T12 region at a current of 20 mA and in the T10-T11 region at a current of 30 mA. At each stimulation site there is a clear threshold for initiating action potentials and the threshold is assumed to be determined largely by the size of the axons. One of the differences in [9] and in the estimates in Table 2 is most likely due to the uncertainty in the size of the neurons and the initiating voltage for the action potential. Indeed, since after the action potential starts, the processes related to the opening of ion channels are dominant, while the influence of external currents charging the membrane of axons ceases to play a role. It should also be noted that the recovery time of the action potential is on the order of 0.01 s, which is much less than the time between the current pulses (Fig. 1).

It should be noted that the conductivity of media in the spinal nerve region varies over a wide range. So conductivity of epidural fat is 0.04 S/m, cerebrospinal fluid is 1.7 S/m, white mater (transversal) is 0.083 S/m, white mater (longitudinal) is 0.6 S/m, grey matter is 0.23 S/m [15,24].

In this model, we assumed that the spinal nerve axons are located far enough apart, so that they can be treated independently. Since the electric field from a charged cylinder falls as $a^2/r^2$ (13), then the interaction of neurons with each other can be neglected if the distance between them is larger than the diameter of the axon.

**Table 2.** Estimations of the excitation of the action potential threshold for various types of neurons (Eq.14) at $\sigma = 0.061 \, \Omega^{-1} m^{-1}$, $U_m = 10$ mv, $R_{el} = 0.9 \cdot 10^{-2}$ m.

| Type | Radius [20] [$\mu m$] | Average [20] [$\mu m$] | Threshold [mA] |
|---|---|---|---|
| α | 6-10 | 8 | 9.7 |
| β | 2.5-6 | 4.25 | 20 |
| γ | 1.5-3 | 2.25 | 38 |
| δ | 1-2.5 | 1.75 | 50 |

In accordance with (13), at a fixed current in the electrolyte, the potential difference on the membrane is directly proportional to the radius of the neuron and is inversely proportional to the conductivity of the electrolyte. The thresholds of excitation of the action potentials in the lumbosacral regions of the spinal cord observed in [9] at currents of ≈10mA, 20mA and 30mA can be attributed largely to differences in the radii $\alpha, \beta, \gamma$ - motor neurons with the site of stimulation at three different spinal levels. This fact can explain the selectivity of the action of impulse current on the excitation of the activity of various neurons at each stimulation site.

The angular dependence of the potential difference on the membrane (14) does not play a special role, because for the excitation of the action potential, it is sufficient to have a local overvoltage on the axon membrane so that the action potential from the excitation region begins to propagate both along and across the nerve fiber. A section of the initial segment of the axon of radius $R_{ax}$, in which the AP is excited, is shown in Fig. 7. The red spot marks the region in which the potential difference between the inner and outer surfaces of the membrane exceeds the threshold of excitation of the action potential, $(\sigma_{out} - \sigma_{in}) \cdot \frac{d_m}{\epsilon_0 \varepsilon_m} > U_t$ ; $\sigma_{out}$, $\sigma_{in}$ are the surface charge densities on the outer and inner sides of the membrane, $d_m$ is the membrane thickness, $\epsilon_0, \epsilon_m$ are the dielectric constant of vacuum and the relative permittivity of the membrane, correspondingly.

The arrows indicate the direction of propagation of the action potential. Obviously, after a while $t_0 > 2\pi R_{ax}/v_{ap}$, where $v_{ap}$ is speed of propagation of action potential, it is possible to neglect the angular dependence, that is, the action potential has a cylindrical symmetry.

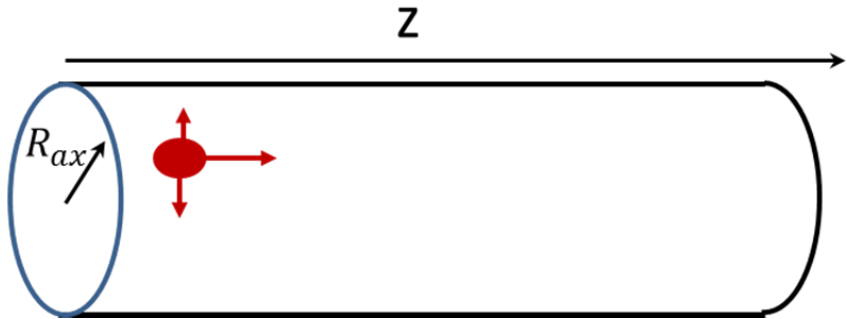

**Figure 7.** A section of the initial segment. $R_{ax}$ is the radius of the axon. The red spot corresponds to the region in which the potential difference between the inner and outer surfaces of the membrane exceeds the threshold of excitation of the action potential. The arrows indicate the direction of propagation of the action potential.

The model considered here assumes that the distance between neurons exceeds 1-2 diameters of the neuron. If the distance between the neurons is less than the diameter, then they cannot be considered as single, i.e. which are in a constant field of currents in the intercellular electrolyte. The refined model for an arbitrary distance between neurons will be studied in future.

The above threshold excitation mechanism of action potentials in motor neurons is inherently close to the ephaptic coupling effect discussed in [19]. The difference is that at ephaptic coupling, currents in the extracellular fluid arise due to the propagation of an action potential in activated neurons, whereas in the present work currents in a conducting intercellular medium are induced by an external source. It was shown in [19] that for a sufficiently close proximity of axons, the action potential propagating in one axon can initiate an action potential in a nearby axon. This is due to the currents arising in the electrolyte in the vicinity of the active axon, charging the membranes of the neighboring axon to a potential sufficient to stimulate the action potential. This is the essence of the ephaptic coupling between neurons in the neural tissue [26,27]. In our opinion, in order to excite the action potential in a damaged fiber, the action potential starts in one axon, for example, on a periphery or on a less damaged section, and, due to ephaptic coupling, spread to the neurons of entire damaged section of the fiber.

The mechanism of excitation of the action potential by external electrodes, in combination with the ephaptic coupling mechanism proposed in [19], will allow us to construct an adequate mathematical model corresponding to the experiments.

## V. Concluding remarks

It is shown that the mechanism of action potential excitation by the currents flowing in the intercellular conducting medium, proposed in [19,20], could explain the results of experiments, in which spontaneous limb motion and excitation of action potentials of multiple motor pools can be produced from the application of stimulating current pulses over uninjured and injured regions of the spinal cord [5-9].

The proposed model can be used effectively to predict and optimize the stimulation of nervous activity by external sources of current.

## Acknowledgements
The authors are grateful to Yu.P. Gerasimenko and V.R. Edgerton for bringing attention to their works and valuable discussions.